\documentclass[aps,prb,twocolumn,showpacs,superscriptaddress]{revtex4}
\usepackage{graphicx}
\usepackage{amssymb,amsmath}
\usepackage{color}

\begin{document}

\title{Phase Coherent Transport in Graphene Nanoribbons and \newline Graphene Nanoribbon Arrays}

\author{S.~Minke}
\affiliation{Institute of Experimental and Applied Physics, University of Regensburg, 93040 Regensburg, Germany}

\author{J.~Bundesmann}
\affiliation{Institute of Theoretical Physics, University of Regensburg, 93040 Regensburg, Germany}

\author{D.~Weiss}
\affiliation{Institute of Experimental and Applied Physics, University of Regensburg, 93040 Regensburg, Germany}

\author{J.~Eroms}
\email{jonathan.eroms@physik.uni-regensburg.de}
\affiliation{Institute of Experimental and Applied Physics, University of Regensburg, 93040 Regensburg, Germany}

\date{\today}

\begin{abstract}
We have experimentally investigated quantum interference corrections to the conductivity of graphene nanoribbons at temperatures down to 20 mK studying both weak localization (WL) and universal conductance fluctuations (UCF). Since in individual nanoribbons at millikelvin temperatures the UCFs strongly mask the weak localization feature we employ both gate averaging and ensemble averaging to suppress the UCFs. This allows us to extract the phase coherence length from both WL and UCF at all temperatures. Above 1 K, the phase coherence length is suppressed due to Nyquist scattering whereas at low temperatures we observe a saturation of the phase coherence length at a few hundred nanometers, which exceeds the ribbon width, but stays below values typically found in bulk graphene. To better describe the experiments at elevated temperatures, we extend the formula for 1D weak localization in graphene, which was derived in the limit of strong intervalley scattering, to include all elastic scattering rates.
\end{abstract}


\maketitle

\section{Introduction}

Phase coherent effects in graphene are determined by the combined action of several scattering mechanisms. In the past, extensive studies have been performed on those effects in bulk graphene~\cite{PhysRevLett.89.266603, Morozov2006,McCann2006,PhysRevLett.97.196804, PhysRevLett.98.136801, Kechedzhi2007, Tikhonenko2008, Ki2008, Tikhonenko2009, Nature446.56, Kharitonov2008, Kechedzhi2008, Kechedzhi2009, Ojeda2010}. Little attention, however, has been paid to phase coherent behavior in graphene nanoribbons (GNRs) where lateral confinement causes a crossover from 2D to 1D behavior and additional scattering is introduced at the edges of the ribbons.

In the experiments of Morozov \textit{et al.}~\cite{Morozov2006} on bulk graphene strong suppression of weak localization was observed. A theoretical description of the phase coherent effects was given by McCann \textit{et al.}~\cite{McCann2006}, where elastic scattering mechanisms (intra- and intervalley scattering)
determine if weak localization (WL), weak antilocalization (WAL) or none of them is observed. If there is neither intravalley scattering nor intervalley scattering weak antilocalization is found, as expected for chiral quasiparticles associated with Berry phase~$\pi$~\cite{PhysRevLett.89.266603}. Intravalley scattering tends to suppress the chiral nature of quasiparticles and, thus, destroys localization, whereas intervalley scattering tends to restore the weak localization effect~\cite{Kechedzhi2007}. In further experiments the phase coherent effects could be interpreted by this theoretical description~\cite{Tikhonenko2008, Ki2008, Tikhonenko2009}. Furthermore it was found that by changing the carrier density and/or the temperature, it was possible to alter the ratio of various scattering rates and observe a transition from WL to WAL as the chiral nature of the charge carriers was restored~\cite{Tikhonenko2009, Pezzini2012}.\\
In the case of graphene nanoribbons, however, due to scattering at the edges, intervalley scattering is predicted to be the most important mechanism leading to the observation of weak localization~\cite{McCann2006}. To our knowledge, up to now there are no extensive experimental studies reported on the analysis of weak localization in GNRs and the theoretical predictions still need to be verified experimentally.\\

\begin{figure}
\begin{center}
\includegraphics[width=8cm]{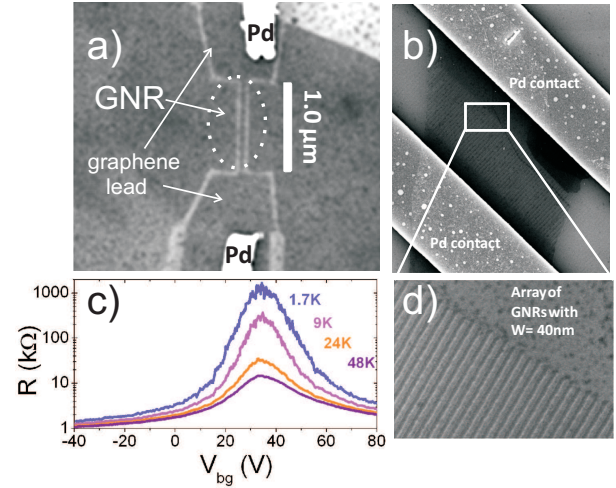}
\end{center}
\caption[GNR]{(a) Scanning electron microscope image of a typical sample from this work. The length of the GNRs is 1~$\mu$m, the width 70~nm. Two palladium contacts are visible. (b)~Array of GNRs with a GNR length of 1~$\mu$m in between two palladium contacts. (c) Sample C: Two-terminal resistance as a function of V$_{bg}$ for different temperatures at zero magnetic field. (d)~Zoom-in of the GNR array, every GNR has a width of 40~nm and a spacing of 30~nm to the next, the zoom-in area of panel~(b) is marked in white.}
\label{GNR1}
\end{figure}
Another correction to the conductivity are universal conductance fluctuations (UCFs), which appear when the sample length does not greatly exceed the phase coherence lengths. In our samples they are clearly visible. In graphene, these fluctuations are sensitive not only to the phase coherence length and the thermal length, but also to elastic scattering (intervalley scattering and intravalley scattering). The conductance variance strongly depends on the exact types of elastic scattering present in the sample and is a factor, $\alpha$, times larger compared to a usual metal because of valley degeneracy. If all scattering effects are negligible $\alpha=4$. For weak intervalley scattering and either strong intravalley scattering or strong trigonal warping $\alpha=2$ and for strong intervalley scattering $\alpha=1$~\cite{Kharitonov2008, Kechedzhi2008}. Experiments on graphene analyzed the universal conductance fluctuations by the correlation function~\cite{Ojeda2010, Kechedzhi2009} and showed that those fluctuations can be used, for example, for thermometry~\cite{Kechedzhi2009}. But up to now no studies on one dimensional graphene structures were performed.\\
Both effects, namely the weak localization as well as the universal conductance fluctuations, allow us to extract the phase coherence length $L_{\varphi}$ in an independent way.
Therefore, we performed experiments on graphene nanoribbons to study both effects.

\section{Experimental details}

Single layer graphene is deposited on a highly doped silicon wafer with a 300~nm thick SiO$_{2}$ layer by conventional exfoliation~\cite{NovoselovPNAS}. The flakes were imaged under an optical microscope and their position was detected with respect to predefined markers. The graphene nanoribbons (GNRs), as well as the arrays of GNRs, were fabricated by electron beam lithography and oxygen plasma reactive ion etching. The ribbon length was 1~$\mu$m and the ribbon width $W$ varies between 40~nm and 80~nm. For the transport measurements palladium contacts were attached to the GNRs, using standard electron beam lithography and thermal evaporation. Micrographs of typical samples are shown in Fig.~\ref{GNR1}.
Electronic characterization and magnetotransport measurements were done in two different cryostats with temperatures ranging from 1.7~K to 125~K and $T=$~20~mK to 900~mK, respectively, with magnetic fields up to $B=$~16~T. The measurements were done in two terminal geometry using standard lock-in technique with frequencies of 13 or 17~Hz and an excitation current of 10~nA at Kelvin temperatures and 0.5~nA at millikelvin temperatures, respectively. To induce charge carriers in GNRs a gate voltage up to $\pm$80~V was applied between the graphene and the Si wafer, see Fig.~\ref{GNR1}(c) for typical backgate measurements. Conductance measurements as a function of backgate voltage, temperature and magnetic field were done on many different devices yielding consistent results. We show representative data for two individual graphene nanoribbons and two graphene nanoribbon arrays. For sample parameters and studied temperature range see Table~\ref{samples}.

\begin{table*}[h]
\centering
\begin{tabular}{c c c c c c c c c c }
\hline
\hline
Sample   && $N$ && $W$(nm) & Temperature & $\mu$ (cm$^{2}/$Vs) & $n$ (10$^{16}$ m$^{-2}$) & $L_{mfp}$ (nm) &  $D$ (m$^{2}/$s)\\
\hline
A && 1 && 40 & 1.7~K - 125~K   & 165 & 5.3 & 4.4 & 0.022\\
A && 1 && 40 & 20~mK - 900~mK  & 165 & 5.3 & 4.4 & 0.022\\
B && 1 && 40 & 20~mK - 900~mK  & 330 & 50.6 & 27.4 & 0.014\\
C && 46 && 40 & 1.7~K - 125~K  & 680 & 5.4 & 18.5 & 0.009\\
D && 23 && 80 & 20~mK - 900~mK & 500 &  4.3 &12.1 & 0.006\\
\hline
\hline
\end{tabular}
\caption{Characteristic parameters of the different samples: Number of ribbons $N$, ribbon width $W$, mobility $\mu$, carrier density $n$, mean-free path $L_{mfp}$ and diffusion constant $D$.}
\label{samples}
\end{table*}

\section{Results and Discussion}

\begin{figure}
\begin{center}
\includegraphics[width=7.5cm]{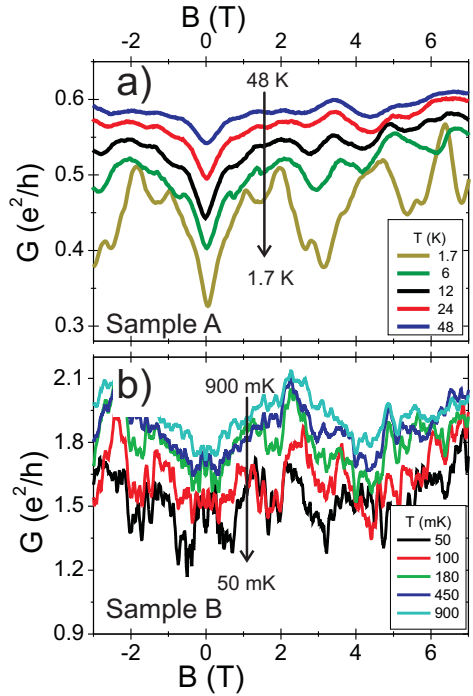}
\end{center}
\caption[Messungen]{(Color online) The conductance $G$ as a function of magnetic field $B$ of a individual GNR shows quantum interference phenomena for temperatures from $T =$ 1.7~K to 48~K (a) and $T =$ 50~mK to 900~mK (b).}
\label{MessungenA}
\end{figure}

Figure~\ref{MessungenA}(a) shows the magnetotransport data collected for a 40~nm wide individual GNR at temperatures from $T =$~1.7~K to 48~K. Weak localization is observed at low fields ($|B|<$1.5~T) as well as universal conductance fluctuations, whose amplitude increases with decreasing temperature.

For mK- temperatures large universal conductance fluctuations overlay the weak localization feature of the individual GNRs~[Fig.~\ref{MessungenA}(b)].
In order to still determine phase coherent properties different methods can be used: (i) Gate averaging: by adding up the measurement traces of different gate values one obtains an average conductance which shows a clear conductance dip and which allows to fit weak localization. (ii) Ensemble averaging: Structuring an array of many graphene nanoribbons in parallel suppresses the UCFs and the phase coherence length can be obtained from fitting the weak localization feature. \\ Furthermore we analyze the universal conductance fluctuations: the phase coherence length can be determined by calculating the autocorrelation function of the UCFs or by analyzing the amplitude of the UCFs.\\ In the following sections we present all the different methods mentioned above and finally we compare the results.

\subsection{Weak localization in individual graphene nanoribbons}\label{ÜA}
\begin{figure}
\centering
\includegraphics[width=6.5cm]{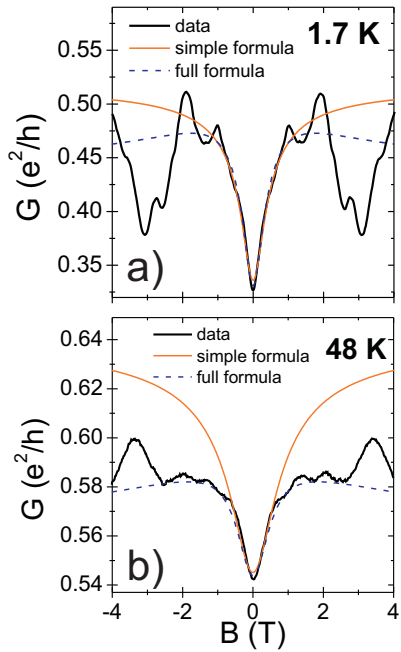}
\caption[Messungen]{(Color online) Sample A: Conductance $G$ as a function of magnetic field $B$ at (a) $T =$~1.7~K and (b) $T =$~48~K. The weak localization feature is fitted by using Eq.~\ref{WLformula1Dxxx} (orange) and Eq.~\ref{WLformula1Dall} (blue dashed line). Both fit formulas reproduce the data well at low temperatures (a), but at higher temperature Eq.~\ref{WLformula1Dall} is more appropriate.}
\label{WLfit3}
\end{figure}

First we analyze and interpret the weak localization effects of single GNRs by fitting to theory~\cite{McCann2006, Beenakker1991}.
The resistivity correction $\delta\rho(B)/\rho^{2}$ is given by the following formula valid in the limit of a very short intervalley scattering time~\cite{McCann2006} and by fitting the magnetotransport data one can determine the phase coherence length $L_{\varphi}$.
\begin{equation}
\label{WLformula1Dxxx}
\frac{\delta\rho(B)}{\rho^{2}}=
\frac{2e^{2}\sqrt{D}}{h} \left( \frac{1}{\tau_{\varphi}}+ \frac{1}{\tau_{B}} \right)^{-1/2} ,
\end{equation}
with the diffusion coefficient $D$, the dephasing time \mbox{$\tau_{\varphi} = L_{\varphi}^{2} D^{-1}$} and the magnetic relaxation time \mbox{$\tau_{B}=3 \hbar^{2}/ (D W^{2}e^{2}B^{2})$}.
The above formula is valid if the magnetic length $L_{m}= \sqrt{\hbar / e B}$ is larger than the ribbon width. For some samples, $L_m$ is on the order of $W$ in the field range considered here. However, since the phase coherence length is extracted from the behavior around $B=0$, the 1D formula is still appropriate to determine $L_{\varphi}$. Also, we fitted the data with both the 1D and 2D formula~\cite{McCann2006} and found that the 2D formula was not able to describe the data well.
Having a closer look at the individual fits, one recognizes that at low temperatures this simple fit formula (Eq.~\ref{WLformula1Dxxx}) reproduces the weak localization feature well at low temperature [Fig.~\ref{WLfit3}(a)]. However at higher temperatures the magnitude of the effect is overestimated [Fig.~\ref{WLfit3}(b)], because the phase coherence length and the intervalley scattering length are of the same order.

Therefore we generalized Eq.~\ref{WLformula1Dxxx} to account for a finite intervalley scattering time by including other relevant elastic scattering times, as done previously~\cite{McCann2006} for two-dimensional graphene. Usually, the WL correction is described in terms of particle-particle correlation functions, so called Cooperons. In two-dimensional graphene $\delta g$ is determined by the interplay of one pseudospin singlet~($C_{0}^{0}$) and three triplet~($C_{0}^{x}, C_{0}^{y}, C_{0}^{z}$) Cooperons, $\delta g \propto -C_{0}^{0}+C_{0}^{z}+C_{0}^{x}+C_{0}^{y}$ and their corresponding relaxation rates~(cf. Ref.~\cite{McCann2006}). For graphene nanoribbons the four Cooperons $C_{0}^{x}, C_{0}^{y}, C_{0}^{z}$ and $C_{0}^{0}$ need to be considered in a similar fashion. Therefore we have to include the contributions from one Cooperon $C_{0}^{z}$ (with~2~$\tau_{i}^{-1}$, where $\tau_{i}^{-1}$ is the intervalley scattering rate) and from two Cooperons $C_{0}^{x}$ and $ C_{0}^{y}$ (with~$\tau_{\ast}^{-1}$, which includes both the inter- and intravalley scattering rates). This leads to the following formula \cite{McCannWL}:
\begin{equation}
\label{WLformula1Dall}
\begin{split}
\frac{\delta\rho(B)}{\rho^{2}}&=
\frac{2e^{2}\sqrt{D}}{h}\Bigg\{ \left( \frac{1}{\tau_{\varphi}}+ \frac{1}{\tau_{B}} \right)^{-1/2} \\
 & -  \left( \frac{1}{\tau_{\varphi}}+ \frac{2}{\tau_{i}}+ \frac{1}{\tau_{B}} \right)^{-1/2}\\
 & -2 \left( \frac{1}{\tau_{\varphi}}+ \frac{1}{\tau_{\ast}}+ \frac{1}{\tau_{B}} \right)^{-1/2}  \Bigg\},
\end{split}
\end{equation}
Here all scattering terms relevant in two-dimensional graphene ($\tau_{\varphi}^{-1}$, $\tau_{B}^{-1}$, $\tau_{i}^{-1}$ and $\tau_{\ast}^{-1}$) are included, with the corresponding lengths \mbox{$L_{\varphi,i,\ast}=\sqrt{D \tau_{\varphi,i,\ast}}$}. Fitting the data with Eq.~\ref{WLformula1Dall}, with the intervalley scattering length $L_{i}$ about the ribbon width and the inter- and intravalley scattering length $L_{\ast}$ about a few nanometers, one obtains much better fits than with Eq.~\ref{WLformula1Dxxx} especially for higher temperatures, cf. Fig.~\ref{WLfit3}. For sample~A a phase coherence length $L_{\varphi}$ between 50~nm and 100~nm can be extracted, Fig.~\ref{results}(a). As it turns out, the phase coherence lengths obtained by Eq.~\ref{WLformula1Dxxx} and Eq.~\ref{WLformula1Dall} are very similar, which proves the robustness of $L_{\varphi}$ and confirms the validity of Eq.~\ref{WLformula1Dall}.\\

\begin{figure}
\begin{center}
\includegraphics[width=8cm]{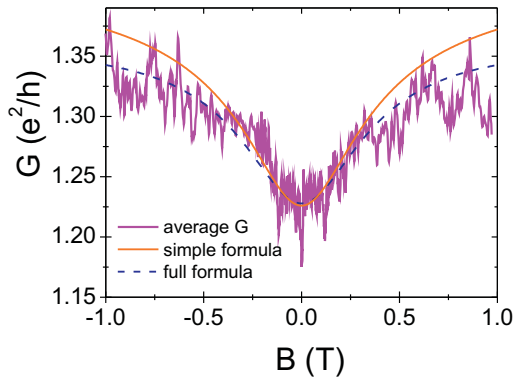}
\end{center}
\caption[GNR]{(Color online) Sample B, gate-averaging: The magnetic field dependence of the conductance was measured at different gate-voltages and the arithmetic mean was calculated. The average conductance $G$ of the 40~nm GNR at $T=$~20~mK clearly shows the weak localization feature. Fitting the conductance dip by Eq.~\ref{WLformula1Dxxx} and Eq.~\ref{WLformula1Dall} yields a phase coherence length of 100~nm.}
\label{GNR}
\end{figure}

Decreasing the temperature, universal conductance fluctuations strongly overlay the WL feature, cf. Fig.~\ref{MessungenA}(b). In order to extract the weak localization and thus the phase coherence length one can do an \textit{averaging over different gate voltages}. For sample B, the magnetotransport was measured for different gate voltages (from $-40$~V to $-20$~V in steps of 1~V) at $T=$~20~mK. The individual traces show strong conductance fluctuations, but by adding up those twenty-one measurements one obtains an average conductance which shows a clear conductance dip, Fig.~\ref{GNR}.
Fitting this feature one obtains a phase coherence length of 100~nm [Fig.~\ref{results}(c), purple star], which is in a reasonable order of magnitude.

We also performed a number of gate dependent experiments on some of the samples but we only saw a decrease of $L_{\varphi}$ around the charge neutrality point, cf. Ref.~\cite{Chen2010}. We never observed a transition to weak antilocalization. Following Tikhonenko  \textit{et al.}~\cite{Tikhonenko2009} this is also not to be expected since they observed the transition in clean bulk graphene when $\tau_{\varphi}$ became shorter that the elastic scattering times. Here, due to nanopatterning,
the elastic scattering times are so short that we could not
reach this regime~\cite{Ortmann2011}.\\

\subsection{Weak localization in arrays of graphene nanoribbons}

\begin{figure}
\begin{center}
\includegraphics[width=7.5cm]{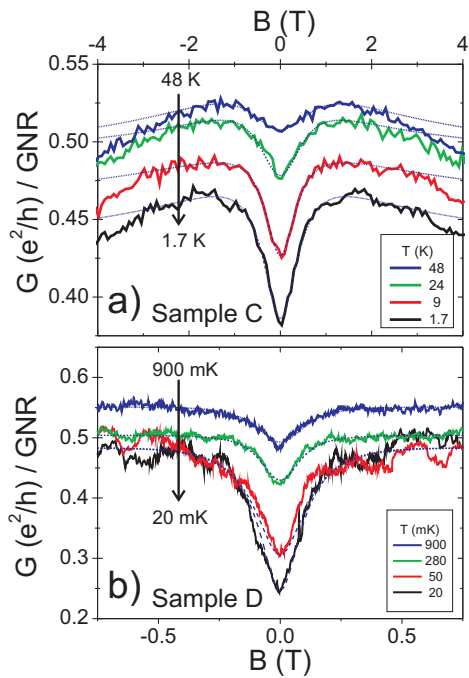}
\end{center}
\caption[Messungen]{(Color online) Magnetotransport data of a GNR array. In comparison to the data of individual ribbons Fig.~\ref{MessungenA}, the measurements of the arrays clearly show a suppression of the universal conductance fluctuations for all temperatures (a) and (b), whereas the weak localization feature is not affected. Blue dashed lines are best-fit curves to Eq.~\ref{WLformula1Dall}.}
\label{MessungenB}
\end{figure}

Also \textit{ensemble averaging} by measuring arrays of graphene nanoribbons suppresses the UCFs. Therefore arrays of graphene nanoribbons were fabricated and the conduction per ribbon was calculated. As expected, the parallel arrangement of the nanoribbons leads to a suppression of the universal conductance fluctuations, whereas weak localization is not suppressed [Fig.~\ref{MessungenB}(a) and (b)]. Thus the phase coherent effects can be separated and the weak localization feature can be fitted again with Eq.~\ref{WLformula1Dxxx} or Eq.~\ref{WLformula1Dall}, which were introduced in section~III~A for individual graphene nanoribbons. \\
Fitting the WL dips for sample C and D [Fig.~\ref{MessungenB}(a) and (b)] to Eq.~\ref{WLformula1Dxxx} and Eq.~\ref{WLformula1Dall}, a phase coherence length $L_{\varphi}$ between 30~nm and 80~nm can be extracted for the array of sample~C and between 80~nm and 170~nm for sample~D, Fig.~\ref{results}(b) and (d). \\

\subsection{Conductance fluctuations}

\begin{figure*}[t]
\begin{center}
\includegraphics[width=14cm]{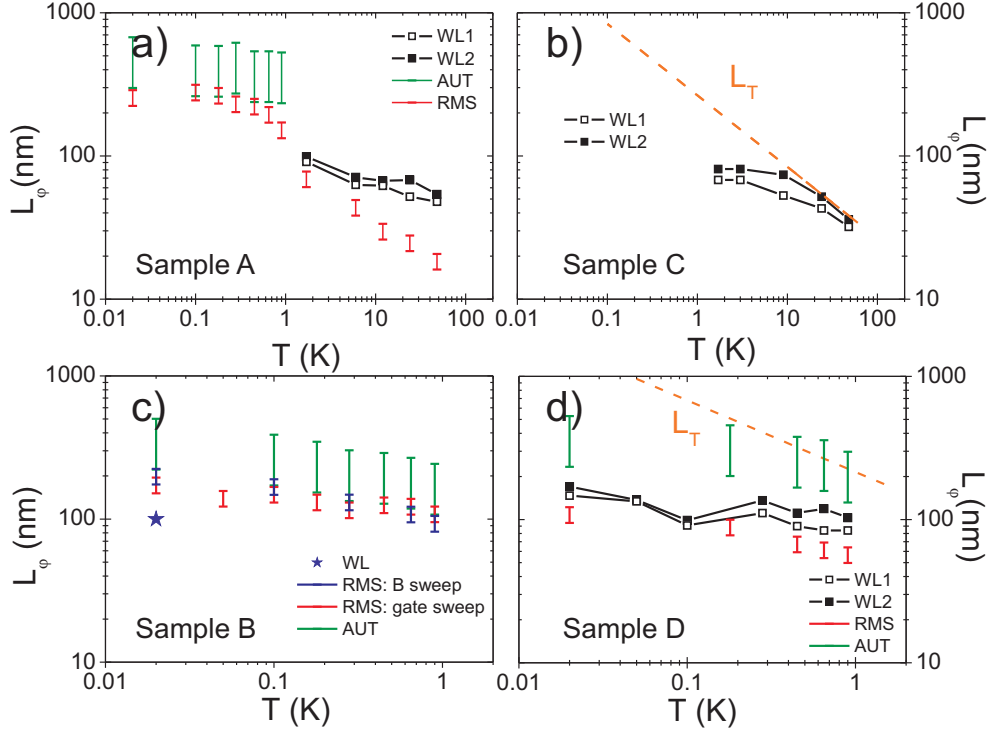}
\end{center}
\caption[results]{(Color) Phase coherence lengths determined via different methods for (a)~sample~A, (b)~sample~C, (c)~sample~B and (d)~sample~D. The phase coherence length $L_{\varphi}$ was determined by different methods like weak localization (WL1 and WL2), the amplitude of the UCFs (RMS) and the autocorrelation function (AUT). The data points obtained by fitting the WL feature to the simple formula (Eq.~\ref{WLformula1Dxxx}) are plotted as black, open squares (WL1) and by fitting the full formula (Eq.~\ref{WLformula1Dall}) as black, filled squares (WL2), respectively.
The orange, dashed lines represent the thermal length $L_{T}$.}
\label{results}
\end{figure*}

Other methods to determine the phase coherence length base upon the analysis of the universal conductance fluctuations.
Therefore let us first interpret the data via the \textit{autocorrelation function}~\cite{Lee1985}. The correlation of the conductance fluctuation is given by the correlation field $B_{C}$ and can be determined from the autocorrelation function defined by $F_{G}(\Delta B)= \int dB \ G(B)\cdot G(B+\Delta B)$, where the integration was done at magnetic field ranges not including the weak localization feature. The correlation function is normalized to the value at $B=$~0~T and the correlation field is defined thus that the function drops to half the maximum value, $F_{G}(B_{c})=0.5 \cdot F_{G}(0)$. 
The phase coherence length can be extracted from the correlation field: $L_{\varphi}=C_{1} \cdot B_{c} \cdot W/\Phi_{0}$, with $C_{1}$ a prefactor between 0.95 for $L_{\varphi} \gg L_{T}$ and 0.42 for $L_{\varphi} \ll L_{T}$~\cite{Beenakker1991} and $\Phi_{0}$ the magnetic flux quantum. We determined the phase coherence length $L_{\varphi}$ for different temperatures giving values between 100~nm and 500~nm for an individual ribbon (sample~B) and between 130~nm and 530~nm for the array of GNRs (sample~D). The values of $L_{\varphi}$ for sample~B and D are summarized in Fig.~\ref{results}(c) and (d), respectively.\\

The \textit{rms amplitude} $\Delta G_{rms}$ of the fluctuations allows us to extract the phase coherence length in an independent way. From the WL measurements we conclude that the $L_{\varphi}$ value is lower than the length of the ribbon $L$. The $L_{\varphi}$ value can be less than or greater than the thermal length $L_{T}=(D\hbar/k_{B}T)^{1/2}$, where $D$ is the diffusion constant. If  $L_{\varphi}<L, L_{T}$ then $\Delta G_{rms}$ depends on $L_{\varphi}$ by the following relation for 1D:

\begin{equation}
\label{SDformula}
\Delta G_{rms}= \alpha \cdot C_{2} \frac{e^{2}}{h}\left(\frac{L_{\varphi}}{L}\right)^{3/2},
\end{equation}
where we set $\alpha=1$ due to strong intervalley scattering that mixes the valleys completely, as already shown by the analysis of the weak localization. In the temperature range considered here, the prefactor $C_2$ ranges from 1.6 to 2.4~\cite{Beenakker1991}.
In Fig.~\ref{results}(c) the temperature dependence of $L_{\varphi}$ extracted from the UCFs obtained by sweeping either the magnetic field (red) or the back gate voltage (blue) is shown for sample~B. In contrast to Ref.~\cite{Bohra},
we do not observe a breakdown of the ergodic hypothesis. Rather, as expected, we find similar fluctuation amplitudes of about 0.4~e$^{2}$/h and thus the values of $L_{\varphi}$ deduced from $\Delta G_{rms} (V_{bg})$ and $\Delta G_{rms} (B)$ match extremely well. Furthermore the absolute values of $L_{\varphi}$ extracted from the rms amplitude $\Delta G_{rms}$ and from the autocorrelation match very well and the temperature dependence is $\sim T^{-0.19}$.\\
For GNR arrays of $N$ ribbons the total conductance is given by \mbox{$G_{N}= N \cdot G_{1}$}, with $G_{1}$ the conductance of a single ribbon. The absolute conductance of the array is $\sim N$ times larger than for an individual GNR. For further analysis, the variance of the conductance is calculated as \mbox{var($G_{N}$)= $N \cdot$ var($G_{1}$)} and the average fluctuation amplitude as \mbox{$\Delta G_{N}=\sqrt{N} \cdot \Delta G_{1}$}. Thus ensemble averaging increases the conductance amplitude only by a factor $\sim \sqrt{N}$ and for the determination of $L_{\varphi}$ by analyzing $\Delta G_{rms}$ of graphene nanoribbon arrays one has to take this factor into account~\cite{Alagha2010}.\\

\subsection{Discussion}

Fitting weak localization with the standard fitting formula for narrow wires (Eq.~\ref{WLformula1Dxxx}) was appropriate only at low temperature, but reaches its limit of applicability at Kelvin temperatures. Therefore we expanded the standard formula to Eq.~\ref{WLformula1Dall}. At high temperature, the corresponding fits describe the measured data much better. At mK-temperatures universal conductance fluctuations mask the WL feature. Different averaging methods (gate- and ensemble-averaging) allow us to still analyze the sample properties. Furthermore the amplitude and the autocorrelation function of the universal conductance fluctuations themselves were analyzed. The determined phase coherence lengths are comparable to the values of $L_{\varphi}$ obtained by fitting the weak localization, whereas the values of $L_{\varphi}$ determined by the autocorrelation function are always slightly higher than those obtained from other methods. We note that there is a discrepancy between the values of $L_{\varphi}$ obtained from the autocorrelation and the rms amplitude of the UCFs for the arrays of nanoribbons [Fig.~\ref{results}(d)]. This may be due to bulk graphene leads, contributing more to the total resistance than for a single GNR, thus making the analysis of the correlation field less reliable.\\

Figure~\ref{results} summarizes the values of the phase coherence lengths determined by different methods. Theoretically the phase coherence lengths is determined from the phase coherence time $\tau_{\varphi}$ by $L_{\varphi}= \sqrt{D \tau_{\varphi}}$. For Nyquist scattering $\tau_{\varphi}$ is proportional to $T^{-2/3}$ and therefore $L_{\varphi}$ is expected to be proportional to $T^{-1/3}$~\cite{Alagha2010, Appenzeller2001, Stojetz2004}. In our experiment, the temperature dependence of $L_{\varphi}$ is about $\sim T^{-0.3}$ at Kelvin temperatures and hence agrees with this model. But for mK-temperature it gets weaker, suggesting a saturating behavior at a few hundred nanometers. Thus $L_{\varphi}$ clearly exceeds the ribbon width for most of our samples, suggesting that the etching process (at the GNR fabrication) does not severely reduce the phase coherent properties of the sample.
However, the values of values of $L_{\varphi}$ in graphene nanoribbons (and graphene antidot lattices~\cite{Jonathan2009}) seem to be smaller than in bulk graphene.
This could in principle be a consequence of the reduced diffusion constant $D$. For our GNR samples A to D we obtain $\tau_{\varphi, A}$=~41~ps, $\tau_{\varphi, B}$=~3~ps, $\tau_{\varphi, C}$=~0.9~ps and $\tau_{\varphi, D}$=~3.8~ps at the lowest temperatures while a bulk graphene sample with $D=0.046$~m$^2$/s showed $L_\varphi=2\ \mu$m  and  $\tau_{\varphi}$=~100~ps at 300 mK. We conclude that  for all nanoribbon samples the phase coherence time as well as the phase coherent length are smaller than in bulk graphene.

The presence of spin flip processes~\cite{Kozikov2012} might explain the lower values of $L_{\varphi}$ in graphene nanoribbons compared to bulk graphene~\cite{Kindermann2009}: Localized spins at the ribbon edges may lead to a de-phasing by spin flip scattering and thus lower the phase coherence length.
Having localized spins at the ribbon edges, one could think of experiments with graphene nanoribbons as spin injectors~\cite{Wimmer2008}.\\

\section{Conclusions}

In conclusion, we have performed magneto-transport measurements in graphene nanoribbons as well as in arrays of GNRs. The observation and analysis of weak localization and universal conductance fluctuations allow us to determine the phase coherent properties of those graphene nanostructures.\\

\begin{acknowledgments}
We would like to thank K. Richter and E. McCann for helpful discussions.
This research was supported by the Deutsche Forschungsgemeinschaft within GRK 1570 (S.M.) and SFB 689 (J.B.).
\end{acknowledgments}


\begin{thebibliography}{29}
\expandafter\ifx\csname natexlab\endcsname\relax\def\natexlab#1{#1}\fi
\expandafter\ifx\csname bibnamefont\endcsname\relax
  \def\bibnamefont#1{#1}\fi
\expandafter\ifx\csname bibfnamefont\endcsname\relax
  \def\bibfnamefont#1{#1}\fi
\expandafter\ifx\csname citenamefont\endcsname\relax
  \def\citenamefont#1{#1}\fi
\expandafter\ifx\csname url\endcsname\relax
  \def\url#1{\texttt{#1}}\fi
\expandafter\ifx\csname urlprefix\endcsname\relax\def\urlprefix{URL }\fi
\providecommand{\bibinfo}[2]{#2}
\providecommand{\eprint}[2][]{\url{#2}}

\bibitem[{\citenamefont{Suzuura and Ando}(2002)}]{PhysRevLett.89.266603}
\bibinfo{author}{\bibfnamefont{H.}~\bibnamefont{Suzuura}} \bibnamefont{and}
  \bibinfo{author}{\bibfnamefont{T.}~\bibnamefont{Ando}},
  \bibinfo{journal}{Phys. Rev. Lett.} \textbf{\bibinfo{volume}{89}},
  \bibinfo{pages}{266603} (\bibinfo{year}{2002}).

\bibitem[{\citenamefont{Morozov et~al.}(2006)\citenamefont{Morozov, Novoselov,
  Katsnelson, Schedin, Ponomarenko, Jiang, and Geim}}]{Morozov2006}
\bibinfo{author}{\bibfnamefont{S.~V.} \bibnamefont{Morozov}},
  \bibinfo{author}{\bibfnamefont{K.~S.} \bibnamefont{Novoselov}},
  \bibinfo{author}{\bibfnamefont{M.~I.} \bibnamefont{Katsnelson}},
  \bibinfo{author}{\bibfnamefont{F.}~\bibnamefont{Schedin}},
  \bibinfo{author}{\bibfnamefont{L.~A.} \bibnamefont{Ponomarenko}},
  \bibinfo{author}{\bibfnamefont{D.}~\bibnamefont{Jiang}}, \bibnamefont{and}
  \bibinfo{author}{\bibfnamefont{A.~K.} \bibnamefont{Geim}},
  \bibinfo{journal}{Phys. Rev. Lett.} \textbf{\bibinfo{volume}{97}},
  \bibinfo{pages}{016801} (\bibinfo{year}{2006}).

\bibitem[{\citenamefont{McCann et~al.}(2006)\citenamefont{McCann, Kechedzhi,
  Fal'ko, Suzuura, Ando, and Altshuler}}]{McCann2006}
\bibinfo{author}{\bibfnamefont{E.}~\bibnamefont{McCann}},
  \bibinfo{author}{\bibfnamefont{K.}~\bibnamefont{Kechedzhi}},
  \bibinfo{author}{\bibfnamefont{V.~I.} \bibnamefont{Fal'ko}},
  \bibinfo{author}{\bibfnamefont{H.}~\bibnamefont{Suzuura}},
  \bibinfo{author}{\bibfnamefont{T.}~\bibnamefont{Ando}}, \bibnamefont{and}
  \bibinfo{author}{\bibfnamefont{B.~L.} \bibnamefont{Altshuler}},
  \bibinfo{journal}{Phys. Rev. Lett.} \textbf{\bibinfo{volume}{97}},
  \bibinfo{pages}{146805} (\bibinfo{year}{2006}).

\bibitem[{\citenamefont{Morpurgo and Guinea}(2006)}]{PhysRevLett.97.196804}
\bibinfo{author}{\bibfnamefont{A.~F.} \bibnamefont{Morpurgo}} \bibnamefont{and}
  \bibinfo{author}{\bibfnamefont{F.}~\bibnamefont{Guinea}},
  \bibinfo{journal}{Phys. Rev. Lett.} \textbf{\bibinfo{volume}{97}},
  \bibinfo{pages}{196804} (\bibinfo{year}{2006}).

\bibitem[{\citenamefont{Wu et~al.}(2007)\citenamefont{Wu, Li, Song, Berger, and
  de~Heer}}]{PhysRevLett.98.136801}
\bibinfo{author}{\bibfnamefont{X.}~\bibnamefont{Wu}},
  \bibinfo{author}{\bibfnamefont{X.}~\bibnamefont{Li}},
  \bibinfo{author}{\bibfnamefont{Z.}~\bibnamefont{Song}},
  \bibinfo{author}{\bibfnamefont{C.}~\bibnamefont{Berger}}, \bibnamefont{and}
  \bibinfo{author}{\bibfnamefont{W.~A.} \bibnamefont{de~Heer}},
  \bibinfo{journal}{Phys. Rev. Lett.} \textbf{\bibinfo{volume}{98}},
  \bibinfo{pages}{136801} (\bibinfo{year}{2007}).

\bibitem[{\citenamefont{Kechedzhi et~al.}(2007)\citenamefont{Kechedzhi, Fal'ko,
  McCann, and Altshuler}}]{Kechedzhi2007}
\bibinfo{author}{\bibfnamefont{K.}~\bibnamefont{Kechedzhi}},
  \bibinfo{author}{\bibfnamefont{V.~I.} \bibnamefont{Fal'ko}},
  \bibinfo{author}{\bibfnamefont{E.}~\bibnamefont{McCann}}, \bibnamefont{and}
  \bibinfo{author}{\bibfnamefont{B.~L.} \bibnamefont{Altshuler}},
  \bibinfo{journal}{Phys. Rev. Lett.} \textbf{\bibinfo{volume}{98}},
  \bibinfo{pages}{176806} (\bibinfo{year}{2007}).

\bibitem[{\citenamefont{Tikhonenko et~al.}(2008)\citenamefont{Tikhonenko,
  Horsell, Gorbachev, and Savchenko}}]{Tikhonenko2008}
\bibinfo{author}{\bibfnamefont{F.~V.} \bibnamefont{Tikhonenko}},
  \bibinfo{author}{\bibfnamefont{D.~W.} \bibnamefont{Horsell}},
  \bibinfo{author}{\bibfnamefont{R.~V.} \bibnamefont{Gorbachev}},
  \bibnamefont{and} \bibinfo{author}{\bibfnamefont{A.~K.}
  \bibnamefont{Savchenko}}, \bibinfo{journal}{Phys. Rev. Lett.}
  \textbf{\bibinfo{volume}{100}}, \bibinfo{pages}{056802}
  (\bibinfo{year}{2008}).

\bibitem[{\citenamefont{Ki et~al.}(2008)\citenamefont{Ki, Jeong, Choi, Lee, and
  Park}}]{Ki2008}
\bibinfo{author}{\bibfnamefont{D.-K.} \bibnamefont{Ki}},
  \bibinfo{author}{\bibfnamefont{D.}~\bibnamefont{Jeong}},
  \bibinfo{author}{\bibfnamefont{J.-H.} \bibnamefont{Choi}},
  \bibinfo{author}{\bibfnamefont{H.-J.} \bibnamefont{Lee}}, \bibnamefont{and}
  \bibinfo{author}{\bibfnamefont{K.-S.} \bibnamefont{Park}},
  \bibinfo{journal}{Phys. Rev. B} \textbf{\bibinfo{volume}{78}},
  \bibinfo{pages}{125409} (\bibinfo{year}{2008}).

\bibitem[{\citenamefont{Tikhonenko et~al.}(2009)\citenamefont{Tikhonenko,
  Kozikov, Savchenko, and Gorbachev}}]{Tikhonenko2009}
\bibinfo{author}{\bibfnamefont{F.~V.} \bibnamefont{Tikhonenko}},
  \bibinfo{author}{\bibfnamefont{A.~A.} \bibnamefont{Kozikov}},
  \bibinfo{author}{\bibfnamefont{A.~K.} \bibnamefont{Savchenko}},
  \bibnamefont{and} \bibinfo{author}{\bibfnamefont{R.~V.}
  \bibnamefont{Gorbachev}}, \bibinfo{journal}{Phys. Rev. Lett.}
  \textbf{\bibinfo{volume}{103}}, \bibinfo{pages}{226801}
  (\bibinfo{year}{2009}).

\bibitem[{\citenamefont{Heersche et~al.}(2007)\citenamefont{Heersche,
  Jarillo-Herrero, Oostinga, Vandersypen, and Morpurgo}}]{Nature446.56}
\bibinfo{author}{\bibfnamefont{H.~B.} \bibnamefont{Heersche}},
  \bibinfo{author}{\bibfnamefont{P.}~\bibnamefont{Jarillo-Herrero}},
  \bibinfo{author}{\bibfnamefont{J.~B.} \bibnamefont{Oostinga}},
  \bibinfo{author}{\bibfnamefont{L.~M.~K.} \bibnamefont{Vandersypen}},
  \bibnamefont{and} \bibinfo{author}{\bibfnamefont{A.~F.}
  \bibnamefont{Morpurgo}}, \bibinfo{journal}{Nature}
  \textbf{\bibinfo{volume}{446}}, \bibinfo{pages}{56} (\bibinfo{year}{2007}).

\bibitem[{\citenamefont{Kharitonov and Efetov}(2008)}]{Kharitonov2008}
\bibinfo{author}{\bibfnamefont{M.~Y.} \bibnamefont{Kharitonov}}
  \bibnamefont{and} \bibinfo{author}{\bibfnamefont{K.~B.}
  \bibnamefont{Efetov}}, \bibinfo{journal}{Phys. Rev. B}
  \textbf{\bibinfo{volume}{78}}, \bibinfo{pages}{033404}
  (\bibinfo{year}{2008}).

\bibitem[{\citenamefont{Kechedzhi et~al.}(2008)\citenamefont{Kechedzhi,
  Kashuba, and Fal'ko}}]{Kechedzhi2008}
\bibinfo{author}{\bibfnamefont{K.}~\bibnamefont{Kechedzhi}},
  \bibinfo{author}{\bibfnamefont{O.}~\bibnamefont{Kashuba}}, \bibnamefont{and}
  \bibinfo{author}{\bibfnamefont{V.~I.} \bibnamefont{Fal'ko}},
  \bibinfo{journal}{Phys. Rev. B} \textbf{\bibinfo{volume}{77}},
  \bibinfo{pages}{193403} (\bibinfo{year}{2008}).

\bibitem[{\citenamefont{Kechedzhi et~al.}(2009)\citenamefont{Kechedzhi,
  Horsell, Tikhonenko, Savchenko, Gorbachev, Lerner, and
  Fal'ko}}]{Kechedzhi2009}
\bibinfo{author}{\bibfnamefont{K.}~\bibnamefont{Kechedzhi}},
  \bibinfo{author}{\bibfnamefont{D.~W.} \bibnamefont{Horsell}},
  \bibinfo{author}{\bibfnamefont{F.~V.} \bibnamefont{Tikhonenko}},
  \bibinfo{author}{\bibfnamefont{A.~K.} \bibnamefont{Savchenko}},
  \bibinfo{author}{\bibfnamefont{R.~V.} \bibnamefont{Gorbachev}},
  \bibinfo{author}{\bibfnamefont{I.~V.} \bibnamefont{Lerner}},
  \bibnamefont{and} \bibinfo{author}{\bibfnamefont{V.~I.}
  \bibnamefont{Fal'ko}}, \bibinfo{journal}{Phys. Rev. Lett.}
  \textbf{\bibinfo{volume}{102}}, \bibinfo{pages}{066801}
  (\bibinfo{year}{2009}).

\bibitem[{\citenamefont{Ojeda-Aristizabal
  et~al.}(2010)\citenamefont{Ojeda-Aristizabal, Monteverde, Weil, Ferrier,
  Gu\'eron, and Bouchiat}}]{Ojeda2010}
\bibinfo{author}{\bibfnamefont{C.}~\bibnamefont{Ojeda-Aristizabal}},
  \bibinfo{author}{\bibfnamefont{M.}~\bibnamefont{Monteverde}},
  \bibinfo{author}{\bibfnamefont{R.}~\bibnamefont{Weil}},
  \bibinfo{author}{\bibfnamefont{M.}~\bibnamefont{Ferrier}},
  \bibinfo{author}{\bibfnamefont{S.}~\bibnamefont{Gu\'eron}}, \bibnamefont{and}
  \bibinfo{author}{\bibfnamefont{H.}~\bibnamefont{Bouchiat}},
  \bibinfo{journal}{Phys. Rev. Lett.} \textbf{\bibinfo{volume}{104}},
  \bibinfo{pages}{186802} (\bibinfo{year}{2010}).

\bibitem[{\citenamefont{Pezzini et~al.}(2012)\citenamefont{Pezzini, Cobaleda,
  Diez, and Bellani}}]{Pezzini2012}
\bibinfo{author}{\bibfnamefont{S.}~\bibnamefont{Pezzini}},
  \bibinfo{author}{\bibfnamefont{C.}~\bibnamefont{Cobaleda}},
  \bibinfo{author}{\bibfnamefont{E.}~\bibnamefont{Diez}}, \bibnamefont{and}
  \bibinfo{author}{\bibfnamefont{V.}~\bibnamefont{Bellani}},
  \bibinfo{journal}{Phys. Rev. B} \textbf{\bibinfo{volume}{85}},
  \bibinfo{pages}{165451} (\bibinfo{year}{2012}).

\bibitem[{\citenamefont{Novoselov et~al.}(2005)\citenamefont{Novoselov, Jiang,
  Schedin, Booth, Khotkevich, Morozov, and Geim}}]{NovoselovPNAS}
\bibinfo{author}{\bibfnamefont{K.~S.} \bibnamefont{Novoselov}},
  \bibinfo{author}{\bibfnamefont{D.}~\bibnamefont{Jiang}},
  \bibinfo{author}{\bibfnamefont{F.}~\bibnamefont{Schedin}},
  \bibinfo{author}{\bibfnamefont{T.~J.} \bibnamefont{Booth}},
  \bibinfo{author}{\bibfnamefont{V.~V.} \bibnamefont{Khotkevich}},
  \bibinfo{author}{\bibfnamefont{S.~V.} \bibnamefont{Morozov}},
  \bibnamefont{and} \bibinfo{author}{\bibfnamefont{A.~K.} \bibnamefont{Geim}},
  \bibinfo{journal}{Proc. Nat. Acad. Sci. USA} \textbf{\bibinfo{volume}{102}},
  \bibinfo{pages}{10451} (\bibinfo{year}{2005}).

\bibitem[{\citenamefont{Beenakker and van Houten}(1991)}]{Beenakker1991}
\bibinfo{author}{\bibfnamefont{C.~W.~J.} \bibnamefont{Beenakker}}
  \bibnamefont{and} \bibinfo{author}{\bibfnamefont{H.}~\bibnamefont{van
  Houten}}, \bibinfo{journal}{Solid State Physics}
  \textbf{\bibinfo{volume}{44}}, \bibinfo{pages}{1} (\bibinfo{year}{1991}).

\bibitem[{McC()}]{McCannWL}
\bibinfo{note}{Thanks to Edward McCann, Lancaster University, for helpful
  communications.}

\bibitem[{\citenamefont{Chen et~al.}(2010)\citenamefont{Chen, Bae, Chialvo,
  Dirks, Bezryadin, and Mason}}]{Chen2010}
\bibinfo{author}{\bibfnamefont{Y.-F.} \bibnamefont{Chen}},
  \bibinfo{author}{\bibfnamefont{M.-H.} \bibnamefont{Bae}},
  \bibinfo{author}{\bibfnamefont{C.}~\bibnamefont{Chialvo}},
  \bibinfo{author}{\bibfnamefont{T.}~\bibnamefont{Dirks}},
  \bibinfo{author}{\bibfnamefont{A.}~\bibnamefont{Bezryadin}},
  \bibnamefont{and} \bibinfo{author}{\bibfnamefont{N.}~\bibnamefont{Mason}},
  \bibinfo{journal}{J. Phys. Cond. Matt} \textbf{\bibinfo{volume}{22}},
  \bibinfo{pages}{205301} (\bibinfo{year}{2010}).

\bibitem[{\citenamefont{Ortmann et~al.}(2011)\citenamefont{Ortmann, Cresti,
  Montambaux, and Roche}}]{Ortmann2011}
\bibinfo{author}{\bibfnamefont{F.}~\bibnamefont{Ortmann}},
  \bibinfo{author}{\bibfnamefont{A.}~\bibnamefont{Cresti}},
  \bibinfo{author}{\bibfnamefont{G.}~\bibnamefont{Montambaux}},
  \bibnamefont{and} \bibinfo{author}{\bibfnamefont{S.}~\bibnamefont{Roche}},
  \bibinfo{journal}{Europhys. Lett.} \textbf{\bibinfo{volume}{94}},
  \bibinfo{pages}{47006} (\bibinfo{year}{2011}).

\bibitem[{\citenamefont{Lee and Stone}(1985)}]{Lee1985}
\bibinfo{author}{\bibfnamefont{P.~A.} \bibnamefont{Lee}} \bibnamefont{and}
  \bibinfo{author}{\bibfnamefont{A.~D.} \bibnamefont{Stone}},
  \bibinfo{journal}{Phys. Rev. Lett.} \textbf{\bibinfo{volume}{55}},
  \bibinfo{pages}{1622} (\bibinfo{year}{1985}).

\bibitem[{\citenamefont{Bohra et~al.}(2012)\citenamefont{Bohra, Somphonsane,
  Aoki, Ochiai, Akis, Ferry, and Bird}}]{Bohra}
\bibinfo{author}{\bibfnamefont{G.}~\bibnamefont{Bohra}},
  \bibinfo{author}{\bibfnamefont{R.}~\bibnamefont{Somphonsane}},
  \bibinfo{author}{\bibfnamefont{N.}~\bibnamefont{Aoki}},
  \bibinfo{author}{\bibfnamefont{Y.}~\bibnamefont{Ochiai}},
  \bibinfo{author}{\bibfnamefont{R.}~\bibnamefont{Akis}},
  \bibinfo{author}{\bibfnamefont{D.~K.} \bibnamefont{Ferry}}, \bibnamefont{and}
  \bibinfo{author}{\bibfnamefont{J.~P.} \bibnamefont{Bird}},
  \bibinfo{journal}{arXiv:1203.6385}  (\bibinfo{year}{2012}).

\bibitem[{\citenamefont{Alagha et~al.}(2010)\citenamefont{Alagha,
  Est\'evez~Hern\'andez, Bl{\"o}mers, Stoica, Calarco, and
  Sch{\"a}pers}}]{Alagha2010}
\bibinfo{author}{\bibfnamefont{S.}~\bibnamefont{Alagha}},
  \bibinfo{author}{\bibfnamefont{S.}~\bibnamefont{Est\'evez~Hern\'andez}},
  \bibinfo{author}{\bibfnamefont{C.}~\bibnamefont{Bl{\"o}mers}},
  \bibinfo{author}{\bibfnamefont{T.}~\bibnamefont{Stoica}},
  \bibinfo{author}{\bibfnamefont{R.}~\bibnamefont{Calarco}}, \bibnamefont{and}
  \bibinfo{author}{\bibfnamefont{T.}~\bibnamefont{Sch{\"a}pers}},
  \bibinfo{journal}{J. Appl. Phys.} \textbf{\bibinfo{volume}{108}},
  \bibinfo{pages}{113704} (\bibinfo{year}{2010}).

\bibitem[{\citenamefont{Appenzeller et~al.}(2001)\citenamefont{Appenzeller,
  Martel, Avouris, Stahl, Hunger, and Lengeler}}]{Appenzeller2001}
\bibinfo{author}{\bibfnamefont{J.}~\bibnamefont{Appenzeller}},
  \bibinfo{author}{\bibfnamefont{R.}~\bibnamefont{Martel}},
  \bibinfo{author}{\bibfnamefont{P.}~\bibnamefont{Avouris}},
  \bibinfo{author}{\bibfnamefont{H.}~\bibnamefont{Stahl}},
  \bibinfo{author}{\bibfnamefont{U.~T.} \bibnamefont{Hunger}},
  \bibnamefont{and} \bibinfo{author}{\bibfnamefont{B.}~\bibnamefont{Lengeler}},
  \bibinfo{journal}{Phys. Rev. B} \textbf{\bibinfo{volume}{64}},
  \bibinfo{pages}{121404} (\bibinfo{year}{2001}).

\bibitem[{\citenamefont{Stojetz et~al.}(2010)\citenamefont{Stojetz, Hagen,
  Hendlmeier, Ljubovi\'c, Forr\'o, and Strunk}}]{Stojetz2004}
\bibinfo{author}{\bibfnamefont{B.}~\bibnamefont{Stojetz}},
  \bibinfo{author}{\bibfnamefont{C.}~\bibnamefont{Hagen}},
  \bibinfo{author}{\bibfnamefont{C.}~\bibnamefont{Hendlmeier}},
  \bibinfo{author}{\bibfnamefont{E.}~\bibnamefont{Ljubovi\'c}},
  \bibinfo{author}{\bibfnamefont{L.}~\bibnamefont{Forr\'o}}, \bibnamefont{and}
  \bibinfo{author}{\bibfnamefont{C.}~\bibnamefont{Strunk}},
  \bibinfo{journal}{J. Appl. Phys.} \textbf{\bibinfo{volume}{108}},
  \bibinfo{pages}{113704} (\bibinfo{year}{2010}).

\bibitem[{\citenamefont{Eroms and Weiss}(2009)}]{Jonathan2009}
\bibinfo{author}{\bibfnamefont{J.}~\bibnamefont{Eroms}} \bibnamefont{and}
  \bibinfo{author}{\bibfnamefont{D.}~\bibnamefont{Weiss}},
  \bibinfo{journal}{New J. Phys.} \textbf{\bibinfo{volume}{11}},
  \bibinfo{pages}{095021} (\bibinfo{year}{2009}).

\bibitem[{\citenamefont{Kozikov et~al.}(2012)\citenamefont{Kozikov, Horsell,
  McCann, and Fal'ko}}]{Kozikov2012}
\bibinfo{author}{\bibfnamefont{A.~A.} \bibnamefont{Kozikov}},
  \bibinfo{author}{\bibfnamefont{D.~W.} \bibnamefont{Horsell}},
  \bibinfo{author}{\bibfnamefont{E.}~\bibnamefont{McCann}}, \bibnamefont{and}
  \bibinfo{author}{\bibfnamefont{V.~I.} \bibnamefont{Fal'ko}},
  \bibinfo{journal}{Phys. Rev. B} \textbf{\bibinfo{volume}{86}},
  \bibinfo{pages}{045436} (\bibinfo{year}{2012}).

\bibitem[{\citenamefont{Vanevi\'c et~al.}(2009)\citenamefont{Vanevi\'c,
  Stojanovi\'c, and Kindermann}}]{Kindermann2009}
\bibinfo{author}{\bibfnamefont{M.}~\bibnamefont{Vanevi\'c}},
  \bibinfo{author}{\bibfnamefont{V.~M.} \bibnamefont{Stojanovi\'c}},
  \bibnamefont{and}
  \bibinfo{author}{\bibfnamefont{M.}~\bibnamefont{Kindermann}},
  \bibinfo{journal}{Phys. Rev. B} \textbf{\bibinfo{volume}{80}},
  \bibinfo{pages}{045410} (\bibinfo{year}{2009}).

\bibitem[{\citenamefont{Wimmer et~al.}(2008)\citenamefont{Wimmer, Adagideli,
  Berber, Tom\'anek, and Richter}}]{Wimmer2008}
\bibinfo{author}{\bibfnamefont{M.}~\bibnamefont{Wimmer}},
  \bibinfo{author}{\bibfnamefont{I.}~\bibnamefont{Adagideli}},
  \bibinfo{author}{\bibfnamefont{S.}~\bibnamefont{Berber}},
  \bibinfo{author}{\bibfnamefont{D.}~\bibnamefont{Tom\'anek}},
  \bibnamefont{and} \bibinfo{author}{\bibfnamefont{K.}~\bibnamefont{Richter}},
  \bibinfo{journal}{Phys. Rev. Lett.} \textbf{\bibinfo{volume}{100}},
  \bibinfo{pages}{177207} (\bibinfo{year}{2008}).

\end{thebibliography}

\end{document}